
\def\chitil{\widetilde\chi}
\def\tauptaum{\tau^+\tau^-}

\def\hp{H^+}
\def\hm{H^-}
\def\hpm{H^{\pm}}
\def\mhpm{m_{\hpm}}

\def\wpm{W^{\pm}}

\def\mt{m_t}
\def\mb{m_b}

\def\hl{h^0}
\def\hh{H^0}
\def\ha{A^0}
\def\mhl{m_{\hl}}
\def\mhh{m_{\hh}}
\def\mha{m_{\ha}}

\def\lam{\lambda}
\def\munichnlc{{\it Proceedings of the Workshop on ``$\epem$ Linear Colliders
at $500\gev$: the Physics Potential''},  publication DESY 92-123A (1992),
ed. P.M. Zerwas,
Feb. 4 - Sept. 3 (1991) --- Munich, Annecy, Hamburg}
\def\saariselka{{\it Proceedings of the 1st International Workshop
on ``Physics and Experiments with Linear $\epem$ Colliders''},
eds. R. Orava, P. Eerola, and M. Nordberg, Saariselka, Finland,
September 9-14, 1992 (World Scientific Publishing, Singapore, 1992)}

\def\gam{\gamma}
\def\lamnew{\Lambda}
\def\cale{{\cal E}}
\def\calo{{\cal O}}
\def\calm{{\cal M}}
\def\cala{{\cal A}}
\def\stop{{\wtilde t}}
\def\mstop{m_{\stop}}
\def\ep{e^+}
\def\em{e^-}

\def\gev{~{\rm GeV}}
\def\tev{~{\rm TeV}}
\def\pbi{~{\rm pb}^{-1}}
\def\fbi{~{\rm fb}^{-1}}

\def\smv{{\it Proceedings of the 1990 DPF Summer Study on
High Energy Physics: ``Research Directions for the Decade''},
editor E. Berger, Snowmass (1990)}
\def\perspectives{{\it Perspectives on Higgs Physics}, ed. G. Kane, World
Scientific Publishing, Singapore (1992)}

\def\prdj#1{{\it Phys. Rev.} {\bf D{#1}}}
\def\npbj#1{{\it Nucl. Phys.} {\bf B{#1}}}
\def\prlj#1{{\it Phys. Rev. Lett.} {\bf {#1}}}
\def\plbj#1{{\it Phys. Lett.} {\bf B{#1}}}

\def\mt{m_t}
\def\wp{W^+}
\def\wm{W^-}
\def\rta{\rightarrow}
\def\tanb{\tan\beta}
\def\sinb{\sin\beta}
\def\cosb{\cos\beta}

\def\lplm{l^+l^-}
\def\cale{{\cal E}}
\def\calo{{\cal O}}
\def\calm{{\cal M}}

\def\hn{h}
\def\hsm{\phi^0}

\def\mhsm{m_{\hsm}}

\def\mw{m_W}
\def\mz{m_Z}
\def\anti{\overline}

\def\ifmath#1{\relax\ifmmode #1\else $#1$\fi}

\def\3quarter{{\textstyle{3 \over 4}}}

\input phyzzx
\Pubnum={$\caps UCD-93-24$}
\date{July, 1993}

\titlepage
\vskip 0.75in
\baselineskip 14pt
\hsize=6in
\vsize=8.5in
\centerline{\bf HIGGS BOSONS AT AN $\bf\epem$ LINEAR COLLIDER:}
\centerline{\bf THEORY AND PHENOMENOLOGY}
\vskip .5in
\centerline{ J.F. Gunion}
\vskip .075in
\centerline{\it Davis Institute for High Energy Physics,
Dept. of Physics, U.C. Davis, Davis, CA 95616}

\vskip .075in
\centerline{\bf Abstract}
\vskip .075in
\centerline{\Tenpoint\baselineskip=12pt
\vbox{\hsize=12.4cm
\noindent I review recent developments in the theory
and phenomenology of Higgs bosons at an $\epem$ linear collider
with $\sqrt s$ of order 500 GeV.
}}

\vskip .15in
\noindent{\bf 1. Introduction}
\vskip .075in

Perhaps the most fundamental mission of a future high energy
$\epem$ collider will be to reveal the nature and source of electroweak
symmetry breaking.  This could prove to be a relatively straightforward
task at an $\epem$ collider in the case of the minimal Standard Model
(MSM). If the single Higgs boson of the model (the $\hsm$) is sufficiently
light, a narrow Higgs resonance will be found, and the
interactions of (longitudinally polarized) $W$ ($W\equiv \wpm,Z$) bosons
will be perturbative at all energies. Adequate machine energy
will be a necessity, but $\sqrt s$ in the range from 500 GeV to 1 TeV should
suffice. If the $W$ boson sector is strongly interacting, a thorough
investigation of all $WW$ scattering channels will be required before one
can hope to fully understand electroweak symmetry breaking, and much higher
machine energy (2 -- 4 TeV) will be required. This review
will focus entirely on the perturbative scenario and on $\sqrt s$
below 1 TeV. \foot{The	strongly-interacting $W$ scenario is reviewed by
T. Han in these proceedings. Technicolor and related scenarios will also
not be considered here.}
\REF\habers{H.E. Haber, in \saariselka, p. 235.} The summary talk
on Higgs physics at the previous meeting in this series,
Ref.~\habers, is a convenient source of background material.

Even in the context of perturbative theories
containing elementary Higgs bosons, the MSM need not be nature's
choice. Many generalizations have been  discussed,\Ref\hhg{For a review see
J.F. Gunion, H.E. Haber, G.L. Kane, and S. Dawson, {\it The Higgs Hunter's
Guide}, Addison-Wesley, Redwood City, CA (1990).}\ including extensions
of the Higgs sector only, extensions of both the gauge and Higgs sectors,
and supersymmetric generalizations of all these types of models.
Supersymmetric generalizations are particularly attractive in the
perturbative context in that they {\it require} the presence of elementary
spin-zero Higgs fields and solve the well-known naturalness
and hierarchy problems. Thus, they provide an enormously attractive
theoretical framework in which elementary Higgs bosons must exist. Further,
in supersymmetric models, there is always one (or more) light Higgs
boson(s)  with coupling(s) to the $WW$ channels such that $WW$ scattering
remains perturbative at all energies.  The most thoroughly investigated
model is the Minimal Supersymmetric Model (MSSM) in which the Higgs sector
contains two Higgs-doublet fields (the minimum number required in the
supersymmetric context), but there is no extension of the gauge or matter
sectors other than the introduction of the supersymmetric partner
states. In contrast to the single Higgs boson of the MSM, there are five
physical Higgs bosons in a two-Higgs-doublet model.  Assuming that there is
no CP violation in the Higgs sector, they are: the $\hl$, the lightest
CP-even mass eigenstate; the $\hh$ the heavier of the two CP-even mass
eigenstates; the $\ha$, the single CP-odd state; and a charged Higgs pair,
$\hpm$. The resulting phenomenology is much richer than that of the MSM.

In this review, I will focus on recent developments in understanding
how to probe the Higgs sector in three representative cases:
the MSM $\hsm$; the SM with an extended Higgs sector containing two
doublet fields, including the possibility of CP violation; and the MSSM. A
mix of theoretical and phenomenological issues will prove relevant. By
demonstrating that we can thoroughly explore these three quite different
cases at a future $\epem$ collider, we will have considerable confidence in
our ability to probe any perturbative theory with elementary Higgs bosons
that nature may have chosen.

\smallskip
\noindent{\bf 2. The Standard Model Higgs Boson}
\smallskip

Let me first review some well-known theoretical `facts'.
In assessing where we should look for the SM $\hsm$, the first
constraint beyond perturbative unitarity of the $WW$ sector derives
from triviality. All current lattice and related investigations
appear to require that $\mhsm\lsim 650$ GeV if the scale of new
physics, $\lamnew$, is to lie above $\mhsm$.\refmark\hhg\  If $\lamnew$ is as
large as $\sim 10^{15}\gev$, then $\mhsm\lsim 175$ GeV is required
(by the renormalization group equations
\foot{Of course, these same renormalization group equations
have difficulty reproducing the low-energy value
of $\sin^2\theta_W$ in the simplest SU(5) grand unification scheme.}) in
order that the theory remain perturbative up to the scale $\lamnew$.
Finally, in order that the quartic coupling of the Higgs sector
not be driven to negative values (implying instability of the potential)
by the large Yukawa coupling associated with the heavy top quark,
it is necessary that $\mhsm$ lie above an $\mt$-- and $\lamnew$--dependent
lower bound. For $\mt=150$ and $\lamnew\sim 10^{15}$ GeV, for instance,
$\mhsm\gsim 100\gev$ is required in the context of perturbatively computed
renormalization group equations. The lower bound decreases with decreasing
$\lamnew$ and/or $\mt$. Nonetheless, it is entirely reasonable that $\mhsm$
should lie in a range that is somewhat above the current upper limit of
$\sim 60\gev$ set by LEP-I, and quite possibly the $\hsm$ will turn out
to be too heavy to be found at LEP-II (which will probe up to $\mhsm\sim
80-90\gev$ for $\sqrt s\sim 190-200\gev$).  In such a case, an $\epem$
collider of moderate energy would be the ideal machine for detecting
the $\hsm$.

Although the mass of the SM Higgs boson is not known, its couplings
to gauge bosons and fermions are completely determined.  Thus,
the branching ratios and production rates for the $\hsm$ can
be computed as a function of $\mhsm$.  For our immediate purposes,
it is only necessary to recall that for $\mhsm\lsim 150\gev$
the $\hsm$ decays primarily to $b\anti b$, while for $\mhsm\gsim 150\gev$
the $\wp\wm$ and $ZZ$ decay modes become dominant. (Below the
$\wp\wm$ or $ZZ$ threshold,
one of the two $W$'s or $Z$'s must be off-shell, but the branching
fraction can still be substantial.) At $\sqrt s=500\gev$, the two main
production mechanisms are $\epem\rta Z\hsm$ and $\epem\rta \nu\anti\nu
\hsm$ (where the $\hsm$ arises from the fusion of $W$'s emitted from
the $\ep$ and $\em$). The former is dominant for $\mhsm\gsim 160\gev$,
while the latter dominates for lower masses.  Both detection modes
have been thoroughly studied.  In the case of the $Z\hsm$ mode,
\REF\gwhs{P. Grosse-Wiesemann, D. Haidt, and H.J. Schreiber, in
\munichnlc, p. 37.}
\REF\janotstudy{P. Janot, in
\munichnlc, p. 107; preprint LAL-92-27, to appear in Proceedings
of the XXVIIth Rencontres de Moriond, {\it Electroweak Interactions and
Unified Gauge Theories}, Les Arcs, France, March 15-22, 1992.}
\REF\dhkmz{A. Djouadi, D. Haidt, B.A. Kniehl, B. Mele and P.M. Zerwas,
in \munichnlc, p. 11.}
the most recent studies are those  of Refs.~\gwhs-\dhkmz.
A review of these results and further refinements are given in the
contribution by K. Kawagoe.  The conclusion is that at $\sqrt s=500\gev$, the
$\hsm$ can be discovered up to $\mhsm\sim 350\gev$ using the recoil missing
mass technique for events in which the $Z$ decays to $\lplm$ or a tagged
$b\anti b$ pair. (At $\mhsm=350\gev$, Ref.~\gwhs\ obtains
a net signal event rate of
$S=23$ compared to a net background rate of $B=15$ for an integrated
luminosity of $L=50\fbi$.) The $\nu\anti\nu\hsm$ mode has also been
recently reexamined \REF\bargeretal{V. Barger, K. Cheung, B. Kniehl, and
R.J.N. Phillips, \prdj{46} (1992) 3725.} in Ref.~\bargeretal, with the
conclusion that it is possible to discover the $\hsm$ via this production
process for $\mhsm\lsim 300\gev$ using the decay mode $\hsm\rta \wp\wm\rta
4~{\rm jets}$.  (At $\mhsm=300\gev$, $S=30$ and $B=50$ for
$L=50\fbi$.) This conclusion is substantially consistent with earlier studies.
\REF\jgtn{J.F. Gunion and A. Tofighi-Niaki, \prdj{36} (1987) 2671;
\prdj{38} (1988) 1433.}
\REF\hkm{K. Hagiwara, J. Kanzaki, H. Murayama, preprint DTP-91-18 (1991).}
\REF\bb{D. Burke, P. Burchat and A. Petersen,
\prdj{39} (1989) 3515, \prdj{38} (1988) 2735.}
\refmark{\jgtn-\bb}\  A rough summary and
extrapolation to other energies is that $\hsm$ discovery will
be possible up to $\mhsm\sim 0.7\sqrt s$.

It should be noted that rates for both of the above production modes are
determined only by the $\hsm WW$ couplings.  Nonetheless, sensitivity to
the $\hsm f\anti f$ (where $f$ is a fermion) couplings through the $\hsm$
branching ratios is significant if $b$ tagging is available,
except in the case of $f=t$. This will be
discussed shortly. Regarding the $t\anti t \hsm$ coupling,
\REF\gaemersgounaris{K.J.F. Gaemers and G.J. Gounaris, \plbj{77} (1978)
379.}
\REF\djouadittbar{A. Djouadi, J. Kalinowski, and P.M. Zerwas,{\it Mod.
Phys. Lett.} {\bf A7} (1992) 1765; {\it Z. Phys.} {\bf C54} (1992) 255.}
Refs.~\gaemersgounaris\ and \djouadittbar\ claim
that the $\hsm$ will be visible in $\epem\rta t\anti t\hsm$ production for
$\mhsm\lsim 120\gev$ for $L=20\fbi$, and probably to somewhat higher mass
at $L=50\fbi$. This production process would then allow a first
determination of the $t\anti t\hsm$ coupling for a light $\hsm$.

\FIG\widthratios{}
\topinsert
\vbox{\phantom{0}\vskip 5.0in
\phantom{0}
\vskip .5in
\hskip -10pt
\special{ insert scr:hawaii_widthratios.ps}
\vskip -1.65in }
\centerline{\vbox{\hsize=12.4cm
\Tenpoint
\baselineskip=12pt
\noindent
Figure~\widthratios: The ratio of $\Gamma({\rm Higgs}\rta\gam\gam)$
computed for two different model choices for a number of cases.
In the case of the $\hsm$, the
ratio of the width predicted in the presence of an extra heavy generation
to that obtained in the MSM is shown.
For the $\hl$, the ratio
$\Gamma(\hl\rta\gam\gam)/\Gamma(\hsm\rta\gam\gam)$ as a function of
$\mhl=\mhsm$ with $\mha=400\gev$ is plotted.
Squarks and charginos have been taken
to be as light as possible without being observable at the $\sqrt
s=500\gev$ collider.  For the $\hh$ two curves are shown.  The
dot-dashed curve is $\Gamma(\hh\rta\gam\gam)$ in a model with light
charginos ($M=-\mu=150\gev$ in the notation of Ref.~\hhg) divided by the
corresponding width with heavy charginos ($M=-\mu=1\tev$), keeping the
squarks and sleptons heavy (with masses of order 1 TeV).
The dashed curve is $\Gamma(\hh\rta\gam\gam)$ in a model with light
squarks and sleptons  (given by a common soft-SUSY breaking diagonal
mass of $150\gev$ for all squarks and sleptons, with all
off-diagonal mass terms set to zero) divided by the corresponding width
computed with heavy squarks and sleptons, keeping the charginos heavy (as
specified above). For the latter two curves, the ratio of widths is plotted
as a function of $\mhh$ for $\tanb=2$. The top quark mass is taken equal
to 150 GeV for all calculations.
\REF\ghgamgam{J.F. Gunion and H.E. Haber, preprint UCD-92-22 (1992);
and \smv, p. 206.}
This figure is taken from Ref.~\ghgamgam.}}
\endinsert

In the last few years the possibility of employing collisions of
back-scattered laser beams to discover the SM Higgs boson at a linear
$\epem$ collider has been explored.
\REF\barklow{T. Barklow, \smv, p. 440.}
\REF\bbc{D. Borden, D. Bauer, and D. Caldwell, preprint SLAC-PUB-5715
(1992).} \refmark{\ghgamgam-\bbc}\
The event rate is directly proportional to $\Gamma(\hsm\rta\gam\gam)$.
The interest in this mode derives primarily from two facts.
First, observation of the $\hsm$ in this production mode provides
probably the only access to the $\hsm\gam\gam$ coupling at an $\epem$
linear collider.  The $\hsm\rta\gam\gam$ decay channel has (at best) a
branching ratio of order $2\times 10^{-3}$; too few events will
be available in direct $\epem$ collisions to allow detection of such decays.
Second, in principle it might be possible to detect the $\hsm$
for $\mhsm$ somewhat nearer to $\sqrt s$ than the $0.7\sqrt s$ that appears
to be feasible via direct $\epem$ collisions. Indeed, the full $\gam\gam$
center-of-mass energy, $W_{\gam\gam}$, goes into creating the $\hsm$, and
the back-scattered laser beam facility can be configured so that
the $W_{\gam\gam}$ spectrum peaks slightly above $0.8\sqrt s$.

The importance of determining the $\hsm\gam\gam$ coupling derives from
the fact that it is determined by the sum over all 1-loop diagrams
containing any charged particle whose mass arises from the Higgs
field vacuum expectation value.  In particular, the 1-loop contribution
of a charged particle with mass $\gsim\mhsm/2$, approaches a constant value
that depends upon whether it is spin-0, spin-1/2, or spin-1. (The
contributions are in the ratio  $-1/3$ : $-4/3$ : 7, respectively.)
For a light Higgs boson, in the MSM the
dominant contribution is the $W$-loop diagram.  The next most important
contribution is that from the top quark loop, which tends to cancel part of
the $W$-loop contribution.  A fourth fermion generation with both a heavy
lepton, $L$, and a heavy $(U,D)$ quark doublet would lead to still further
cancellation. For $\mhsm\gsim 2\mw$, the $W$-loop contribution decreases,
and the heavy family ultimately dominates. To
illustrate, we show in Fig.~\widthratios\ the ratio of
$\Gamma(\hsm\rta\gam\gam)$ as computed in the MSM (with $\mt=150\gev$) to
that computed in the presence of an extra generation with $m_L=300\gev$ and
$m_U=m_D=500\gev$. Except for $\mhsm$ in the vicinity of $300\gev$,
where the full set (mainly the heavy generation) of contributions
accidentally matches the MSM result, even a rough measurement (or bound) on
the $\hsm\gam\gam$ coupling would reveal the presence of the otherwise
unobservable heavy generation. It is especially interesting to note that a
heavy generation would greatly enhance the event rate (and hence prospects)
for detecting a Higgs boson with mass up near $\sqrt s=500\gev$.

Because of the dominance of the $W$ loop contribution in the three family
case, the $\hsm\gam\gam$ coupling is also very sensitive to any deviations
of the $WW\gam$ and $WW\hsm$ couplings from SM values.
\REF\derujula{A. de Rujula \etal, \npbj{384} (1992) 3.}
\REF\perezt{M.A. Perez and J.J. Toscano, \plbj{289} (1992) 381.}
\refmark{\derujula,\perezt}\  The sensitivity to
anomalies in these couplings can be substantially greater than that
provided by LEP-I data.

How high in mass can the $\hsm$ be detected in $\gam\gam$ collisions
in the case of the MSM?  For $\sqrt s=500\gev$, the range of
interest is that which cannot be accessed by
direct $\epem$ collisions, \ie\ $\mhsm\gsim 350\gev$. In this mass region,
$\hsm\rta ZZ$ decays provide the best signal. Certainly, the tree-level
$\gam\gam\rta\wp\wm$ continuum background completely overwhelms
the $\hsm\rta\wp\wm$ mode. As summarized in Ref.~\ghgamgam,
if there were no continuum $ZZ$ background, and if one of the
$Z$'s is required to decay to $\lplm$, the event rate would be adequate
for $\hsm$ detection up to $\mhsm\sim 400\gev$, \ie\ $\mhsm\sim 0.8\sqrt
s$. Unfortunately, even though there is no
tree-level $ZZ$ continuum background, such a background does arise
at one-loop. A full calculation of this background was performed in
\REF\jikia{G.V. Jikia, \plbj{298} (1993) 224. See also Jikia's
contribution to these proceedings.}
Ref.~\jikia. The $\wpm$ loop is dominant, and leads to a large
rate for $ZZ$ pairs with large mass, when one or both of the $Z$'s
is transversely polarized. This background is such that $\hsm$
observation in the $ZZ$ mode is probably not possible for $\mhsm\gsim
350\gev$, \ie\ no better than what can be achieved in direct
$\epem$ collisions.
\REF\berger{M.S. Berger, preprint MAD/PH/771 (1993).}
This result has been confirmed in the recent independent
calculation of Ref.~\berger.

In general, although the $\gam\gam$ mode may not
extend the discovery reach of an $\epem$ collider, it {\it will} allow
a first measurement of the $\hsm\gam\gam$ coupling of any Higgs
boson that is found in direct $\epem$ collisions. The accuracy that can be
expected has been studied in
\REF\bbcnew{D. Bauer, D. Borden, and D. Caldwell, preprint UCSB-HEP-93-01
(1993).}
Ref.~\bbcnew. They consider two final states: the
$\hsm\rta b\anti b$ channel with $b$-tagging, and the $\hsm\rta ZZ$
channel with one $Z$ required to decay to $\lplm$. In the former case, it is
important, as noted in Ref.~\ghgamgam, to polarize the laser beams so that
the colliding photons have $<\lam_1\lam_2>$ near 1.  This suppresses the
$\gam\gam\rta b\anti b$ background which is proportional to
$1-<\lam_1\lam_2>$. They find that
if $35\lsim\mhsm\lsim 150\gev$, then the $b\anti b$ mode will allow a 5-10\%
determination of $\Gamma(\hsm\rta\gam\gam)$, while for
$185\lsim\mhsm\lsim300\gev$ the $ZZ$ mode will provide a 8-11\%
determination.  In the $150-185\gev$ window, the $WW$ and $b\anti b$ decays
are in competition, and the accuracy of the measurement might not be better
than 20\%.

Other recent developments in $\hsm$ physics reported at this conference
are in five areas. 1) Measuring details of the $\hsm$ couplings and
verifying its quantum numbers, once the $\hsm$ has been discovered.
2) Computation of radiative corrections to various production processes
and decays. 3) Complications (one of which is mentioned above)
in the $\gam\gam\rta\hsm$ discovery mode. 4) Associated production of the
$\hsm$ and other particles in $\gam\gam$ collisions. 5) $e\gam$ collision
mode possibilities.  I shall make only a few remarks on each. More details
can be found in the plenary talks by P. Janot and D. Borden, and in the
various parallel contributions to these proceedings.

\REF\hildreth{M.D. Hildreth, preprint SLAC-PUB-6036 (1993);
and contribution to these proceedings.} The value of $b$ tagging for
separating different decay channels, $X$, of the $\hsm$, and thereby
determining the product $\sigma_{\rm tot}\times BR(\hsm\rta X)$ was
demonstrated in Ref.~\hildreth. The channels $X=b\anti b$, $WW^*$, $c\anti
c+gg$ and $\tauptaum$ yield different numbers of displaced vertices
on average.  By fitting the distribution in the number of displaced
vertices, approximate determinations of $\sigma\times BR$
are possible.  For $\mhsm\sim 140\gev$, the accuracies achieved in the four
channels listed above are roughly 12\%, 24\%, 116\% and 22\%, respectively.
Whether or not this is a useful level, is somewhat model
and situation dependent.  For
instance, in the model where the Higgs sector of the SM is extended to
two-doublets,  the couplings of any one of the three neutral Higgs bosons
to  the various channels can be very different from couplings for the
$\hsm$.  If only a single neutral Higgs boson is observed,
an experimental demonstration that the couplings to $b\anti b$,
$\tauptaum$ and $WW^*$ are within 10-20\% of the values expected for
the $\hsm$ would strongly suggest that the only two-doublet extensions
that should be considered are ones in which one of the Higgs bosons
automatically has couplings that are close to MSM values.  The MSSM
supersymmetric two-doublet extension is a case in point.  There,
the $\hl$ is predicted to have couplings that are not dramatically
different from those predicted for the $\hsm$.  Relatively precise
determinations could be required to discriminate between the $\hsm$
and the $\hl$.

Determination of the quantum numbers of a neutral Higgs boson has been
another active area. I will defer this discussion to the section
on extended Higgs sectors, since it is only by comparison to a model
in which there are Higgs bosons with quantum numbers other than those
of the $\hsm$ that one can assess how useful the  proposed techniques
are.  However, I will state my conclusions here. While various production
and decay distributions are sensitive to whether the Higgs boson is CP-even
or CP-odd, with two exceptions
the production or decay processes considered to date  will only be
visible if the Higgs boson is CP-even, or has a significant CP-even
component. In this case, the distributions will be entirely dominated
by the CP-even prediction and there will be very little sensitivity
to any CP-odd component that might be present. An especially interesting
exception to this generality occurs in considering polarization asymmetries
in $\gam\gam\rta$~Higgs production using the fact that  back-scattered laser
beam photons can be given large polarizations.

I will not dwell on the recent progress made in the
area of radiative corrections.  Details can be found in some
of the parallel contributions.
\Ref\kniehl{See the contributions to these proceedings by
R. Hempfling, B. Kniehl and Y. Kurihara, and references therein.}\
Large corrections to $\hsm\rta b\anti b$ (due primarily to
the running QCD mass of the bottom quark) were found early on
and lead to $\Gamma(\hsm\rta b\anti b)^{\rm 1-loop}\sim 0.55
\Gamma(\hsm\rta b\anti b)^{\rm tree}$ once $\mhsm\gsim 60\gev$.
(See Ref.~\hhg, and references therein.) QCD corrections enhance
the one-loop width for $\hsm\rta gg$ by about 60\%,
\REF\dawson{S. Dawson, \npbj{359} (1991) 283.}
\REF\djouadigg{A. Djouadi, M. Spira, P.M. Zerwas, \plbj{264} (1991) 440.}
\refmark{\dawson,\djouadigg}
whereas corrections to $\Gamma(\hsm\rta\gam\gam)$ are small if
$\mhsm<2\mt$
\Ref\djouadigamgam{A. Djouadi, M. Spira, J.J. van der Bij, and P.M. Zerwas,
\plbj{257} (1991) 187.}
and also generally small for $\mhsm>2\mt$ except for $\mhsm$ such that
there is a large cancellation between the $W$ and $t$ loops.
\REF\melnikov{K. Melnikov and O. Yakovlev, preprint BUDKERINP-93-04
(1993).}
\REF\dszgamgamnew{A. Djouadi, M. Spira, P.M. Zerwas, preprint DESY-92-170
(1993).}
\refmark{\melnikov,\dszgamgamnew}\ Corrections to the $Z\hsm$ production
process are generally small at a $\sqrt s\sim 500\gev$ machine, and, in any
case, are exhaustively studied.
\REF\hempkniehl{J. Fleischer and F. Jegerlehner, \npbj{216} (1983) 469.
B. Kniehl, {\it Z. Phys.} {\bf C55} (1992) 605. R. Hempfling and B. Kniehl,
preprint DESY-93-033 (1993).}
\refmark{\kniehl,\hempkniehl}
\REF\dkk{A. Denner, B.A. Kniehl and J. Kublbeck, in \munichnlc,
p. 31.}
(See also Ref.~\dkk.)
Electroweak corrections to $\hsm$ decays are generally
small.\refmark\kniehl\

A number of contributions to this conference have explored further
backgrounds to detecting the $\hsm$ in the $\gam\gam$ collision mode.
In general, I believe the importance of these additional backgrounds has
been over emphasized, with the exception of the $\gam\gam\rta ZZ$
continuum background from the $W$-loop graphs discussed earlier.
The processes $\gam\gam\rta Z\lplm$ and $Zq\anti q$ yield a reducible
background to the $ZZ$ mode to the extent that the $q\anti q$ or $\lplm$
have mass near $\mz$.
\Ref\bowserchao{D. Bowser-Chao and K. Cheung, preprint NUHEP-TH-92-29
(1993).}\
The magnitude of this background depends upon the detector
resolution. If $<\lam_1\lam_2>$ is not near 1, then this background
can be significant (though not as large as the $ZZ$ continuum
background). However, these processes are proportional to
$1-<\lam_1\lam_2>$ and can be suppressed substantially by appropriate
polarization choices for the incoming back-scattered laser beams.

The $b\anti b$ channel receives a background contribution from
``resolved'' photon processes.
\Ref\ebolietal{O.J.P. Eboli, M.C. Gonzalez-Garcia, F. Halzen and
D. Zeppenfeld, preprint MAD-PH-743 (1993).}\
The most important example is
that where one incoming $\gam$ fragments to a spectator jet and a gluon.
The subprocess $\gam g\rta b\anti b$ then yields a large $b\anti b$ rate.
However, this background will not be a problem in practice for two reasons.
First, it will be possible to veto against the spectator jet that
accompanies the $g$.  This probably already reduces the background to
a level below the true $\gam\gam\rta b\anti b$ continuum.
Second, for the range of $\mhsm$ such that the $b\anti b$ mode is
appropriate ($\mhsm\lsim 150\gev$), the $\hsm$ will already have been
discovered in direct $\epem$ collisions, \ie\  $\mhsm$ will be known.
To study the $\hsm$ in $\gam\gam$ collisions it will be easy to adjust
the machine energy and laser beam polarizations so that the $\gam\gam$
spectrum is peaked at $W_{\gam\gam}\sim\mhsm\sim 0.8\sqrt s$.
(See, for instance, Ref.~\bbcnew.) In this case, the secondary gluon
in the ``resolved''-photon process is quite unlikely to have sufficient
energy to create a $b\anti b$ pair with mass as large as $\mhsm$.

A variety of other final states containing the $\hsm$ can be produced
in $\gam\gam$ collisions.  For instance, the $\gam\gam\rta t\anti t \hsm$
analogue of $\epem\rta t\anti t \hsm$ could provide another measure
of the $t\anti t \hsm$ coupling.
\REF\boostthsm{E. Boos \etal, {\it Z. Phys.} {\bf 56} (1992) 487.}
\REF\keungtthsm{K. Cheung, \prdj{47} (1993) 3750.}
\refmark{\boostthsm,\keungtthsm}
However, because of phase space suppression, the rate for this reaction
is quite small for $\sqrt s=500\gev$, and only becomes competitive
with $\epem\rta t\anti t\hsm$ when $\sqrt s \gsim 1\tev$.
\foot{Radiative corrections to $\gam\gam\rta t\anti t$ and $ZZ$ due to
1-loop Higgs exchange graphs are also sensitive to the $t\anti t\hsm$
coupling. Sufficiently precise measurements of these processes at high
luminosity and energy might allow a determination of the coupling over a
significant range of $\mt$ and $\mhsm$ values, {\it
assuming no other new physics in the 1-loop graphs}.\refmark\boostthsm}

Turning now to $e\gam$ collisions, I merely note here that
$e\gam\rta W\hsm\nu\rta jj b\anti b\nu$ may be viable for $\hsm$
searches for $\sqrt s\gsim 1\tev$.
\REF\hagwz{K. Hagiwara, I. Watanabe, P.M. Zerwas, \plbj{278} (1992) 187.}
\REF\boos{E. Boos, M. Dubinin, V. Ilyin, A. Pukhov, G. Jikia, S.
Sultanov, \plbj{273} (1991) 173.}
\REF\cheung{K. Cheung, preprint NUHEP-TH-93-3 (1993).}
\refmark{\hagwz-\cheung}\
(The last reference includes some background studies.)
This process is interesting in that it probes the $\gam W\rta \hsm W$
subprocess which is determined by a combination of graphs with
different basic SM couplings.  Should the couplings deviate from SM
predictions, the large cancellations among the graphs might
be reduced and the event rate significantly enhanced. Another mode
of interest is $e\gam\rta e\gam\gam\rta e\hsm$, in which a secondary $\gam$
collides with the primary $\gam$ to create the $\hsm$. \Ref\egn{O.J.P.
Eboli, M.C. Gonzalez-Garcia, S.F. Novaes, preprint MAD-PH-753 (1993).}\
The cross section for this process is bigger than that
for $e\gam\rta W\hsm\nu$ and might allow detection of the $\hsm$
at $\sqrt s=500\gev$ in the $b\anti b$ mode. (Resolved photon
backgrounds would have to be suppressed by spectator jet vetoing.)

In summary, it seems clear that the standard
$\epem\rta Z^*\rta Z\hsm$ production mode has as much, or more, potential
for $\hsm$ discovery as any other process studied to date.  Determination of
expected couplings will be possible to a reasonable level, but (as
discussed in the next section) verification  through production or decay
distributions that the $\hsm$ is a {\it pure} CP-even state will not be
easy. We will only know that if we see it in this production mode, it
cannot be pure CP-odd.  Of course, if the cross section level is that
predicted for the maximally coupled $\hsm$ of the MSM, then we can be
rather certain that the observed Higgs boson is either the $\hsm$ or a very
close approximation thereto. (This latter possibility arises naturally in
the MSSM.) $\gam\gam$ and $e\gam$ collisions will primarily be of interest
for determining the $\hsm\rta\gam\gam$ coupling, which potentially
probes new charged particles at high mass scales. Finally, for $\mhsm\lsim
300\gev$, not only can the $\hsm$ be detected in $\gam\gam$ collisions, but
its $\gam\gam$ coupling can be determined to rather good accuracy.

Before closing this MSM section, it is useful to make some comparisons
to hadron supercolliders. I have no space to justify the statements
made regarding hadron colliders. Please refer to
\REF\supercollider{See, for example, the SDC Collaboration, Technical
Design Report, SDC report, SDC-92-201 (1992).}
Refs.~\hhg\ and \supercollider\ for sample studies.
\pointbegin Discovery:  In $\epem$ collisions, discovery will be possible
up to roughly $0.7\sqrt s$, whereas at the SSC/LHC $80\lsim\mhsm\lsim 1\tev$
can be probed.
\point $b\anti b$ and $t\anti t$ couplings: In $\epem$ collisions these
couplings can be determined provided the $b\anti b$
or $t\anti t$  branching ratio of the
$\hsm$ is significant.  At the SSC/LHC, only the
$t\anti t \hsm\rta t\anti t b\anti b$ production/decay mode allows
direct observation of the $\hsm$ in its $b\anti b$ decay channel.
\Ref\dgvi{J. Dai, J.F. Gunion, and R. Vega, preprint UCD-93-18 (1993).}\
This allows sensitivity to the combination $g_{\hsm t\anti t}^2 BR(\hsm\rta
b\anti b)$.  Large deviations from expectations would be apparent
for $\mhsm\lsim 120\gev$, but
precise determinations of the couplings would be difficult.
\point $WW$ couplings: The $\epem\rta Z^*\rta Z\hsm$ recoil mass discovery
mode allows a determination of the $ZZ\hsm$ coupling that is independent
of the Higgs decays.  At hadron colliders, if $\sigma(gg\rta \hsm)$
can be accurately computed, and if acceptances \etc\ of the detector
are well-known, then the $gg\rta\hsm\rta ZZ^*$ or $ZZ$ events allow
a determination of $BR(\hsm\rta ZZ^{(*)})$ to reasonable accuracy.
This, however, does not directly determine the $ZZ\hsm$ coupling.
Extraction of the coupling would have very limited accuracy except
in the $\mhsm$ region between about 130 and 150 GeV, where the
$\hsm$ could also be observed in the $t\anti t b\anti b$ production/decay
mode mentioned above.
\point $\tau\tau$ coupling: At an $\epem$ machine this coupling can be
determined wherever the $\tau\tau$ branching ratio is significant.
At the SSC/LHC, a means for determining the $\hsm\rta\tau\tau$
branching ratio has not been convincingly established.
\point $\gam\gam$ coupling: At an $\epem$ collider, a back-scattered
laser beam facility will allow determination of this coupling over
essentially the same mass range that observation of the $\hsm$
in direct $\epem$ collisions is possible.  At the SSC/LHC, sensitivity
to this coupling will only be significant in the $80\lsim\mhsm\lsim
150\gev$ mass range, where the $\hsm$ can be detected in the
$gg\rta\hsm\rta\gam\gam$ or
$gg\rta t\anti t\hsm\rta t\anti t \gam\gam$ production/decay modes.

In short, so long as the $\epem$ machine energy is sufficient for detection
of the $\hsm$, it will be easier to study its couplings than at the SSC or
LHC.  The primary advantage of the latter machines is their large
discovery reach, which greatly exceeds that available at, for instance,
a $\sqrt s=500\gev$ $\epem$ collider.

Finally, although we have focused on discovery of the
$\hsm$ and measurement of its basic couplings in the simplest reactions,
more complicated final states may ultimately also be of interest.
These include such reactions as $\epem\rta \wp\wm\hsm$, $\gam Z\hsm$,
$ZZ\hsm$, $\nu e \wpm\hsm$, $\nu\anti\nu Z\hsm$ and $\nu\anti \nu
\gam\hsm$. These were studied in
\REF\bhs{V. Barger, T. Han, and A. Stange, \prdj{42} (1990) 777.}
\REF\meleetal{B. Mele, M. Baillargeon, F. Boudjema, F. Cuypers, and
E. Gabrielli; see the contribution by B. Mele to these proceedings.}
Refs.~\bhs\ and \meleetal.  Aside from probing a complicated
combination of SM couplings, these processes could also be sensitive
to anomalous couplings of the $\hsm$ to three gauge bosons.
However, given the current lower bound on $\mhsm$ from LEP, adequate event
rates for such processes will require either large integrated luminosity or
$\sqrt s\gsim 1\tev$.

\smallskip
\noindent{\bf 3. Extended Higgs Sectors}
\smallskip

In this section, I will make some remarks on models in which only
the Higgs sector of the Standard Model is extended.
The most attractive such extension is to two Higgs doublet fields.
For such an extension $\rho=1$ remains automatic at tree
level.\refmark\hhg\ The general two-doublet model is also a convenient
model for exploring the phenomenology of CP violation in the Higgs sector.
If CP is conserved, the two-doublet model yields the five physical Higgs
CP-eigenstates enumerated in the introduction: $\hl$, $\hh$, $\ha$, and
$\hpm$. If the Higgs potential violates CP, then in general all the
neutral states mix, and there are simply three mass eigenstates of mixed CP
character.

Of course higher representations can also be considered, the next most
complicated being a Higgs triplet. As is well-known\refmark\hhg\
$\rho=1$ is not automatic in such a case. Various possibilities for
obtaining $\rho=1$ at tree level can be entertained.  First, it could be
that $\mt$ is very large. In this case the $t-b$ doublet yields a large
positive contribution to $\Delta\rho$ which could by cancelled by the
negative $\Delta \rho$ that would arise from a $T=1$, $Y=\pm2$ complex
triplet representation. Obviously, this would require fine tuning the
triplet vacuum expectation value. A second possibility is to combine one
doublet Higgs field with one real $(T=1,Y=0)$ and one complex $(T=1,Y=2)$
field.\refmark\hhg\ If the neutral members of the two triplet
representations have the same vacuum expectation value, then $\rho=1$ is
maintained at tree level.  In either case, $\rho=1$ is not maintained at
1-loop. Indeed, unlike the case of doublet models $\rho$ is infinitely
renormalized in triplet models (due to the fact that the interactions of
the $B$ gauge field with the Higgs bosons violate custodial SU(2)).
Thus, fine-tuning would be required to maintain $\rho=1$ after
renormalization.  For this reason, these models are generally not in
favor with theorists.  However, this should not stop the experimental
community from searching for the many new signatures that would arise.
At an $\epem$ collider, the most spectacular and characteristic signal
for a triplet model would be the detection of the doubly charged Higgs boson(s)
(using $\epem\rta H^{++}H^{--}$) contained
in complex Higgs triplet representation(s). All that is required
is adequate machine energy, $\sqrt s\gsim 2m_{H^{++}}$.

Returning to the two-doublet model, let me begin by making a few general
remarks.  First, the new CLEO limit on the $b\rta s\gam$ branching ratio
restricts $\mhpm$ to be large,
\REF\hewettbsgam{J. Hewett, \prlj{70} (1993) 1045.}
\REF\bargerbsgam{V. Barger, M.S. Berger, R.J.N. Phillips, \prlj{70} (1993)
1368.}
\refmark{\hewettbsgam-\bargerbsgam}\ in particular,
outside the range (roughly $2\mhpm\lsim 0.8\sqrt s$)
accessible for $\hpm$ detection in $\epem\rta\hp\hm$ at a $\sqrt s=500\gev$
collider.
\Ref\komamiya{S. Komamiya, \prdj{38} (1988) 2158.
P. Eerola and J. Sirkka, \munichnlc, p. 133.
See also the talk at this conference by A. Sopczak.}
(It is important to emphasize that the limits on $\mhpm$ from $b\rta s\gam$
decays obtained in Refs.~\hewettbsgam\ and \bargerbsgam\ assume
no new physics in the loop graphs other than a single charged Higgs boson.)
Second, let us recall that in the CP conserving case the $Z^*\rta Z\hl$
($Z\hh$) cross section is proportional to $\sin^2(\beta-\alpha)$
($\cos^2(\beta-\alpha)$), whereas
the $Z^*\rta \hl\ha$ cross section is proportional to
$\cos^2(\beta-\alpha)$. Here, $\alpha$ is the mixing angle determine by the
$2\times 2$ mass matrix for the CP-even Higgs sector, and $\tanb=v_2/v_1$
with $v_2$ ($v_1$) being the vacuum expectation value of the neutral member
of the Higgs doublet that couples to up (down) quarks.  The
complementarity of the $Z\hl$ and $\hl\ha$ reactions has been pointed out
many times.\refmark\hhg\ Not both reactions can be suppressed by a weak
coupling.  However, it is not impossible that $Z^*\rta Z\hl$
could be kinematically allowed but coupling suppressed, while
$Z^*\rta\hl\ha$ and $Z^*\rta Z\hh$ could be kinematically forbidden.  Then,
no Higgs boson would be seen without raising the machine energy.  In a
general two-doublet model the masses of the Higgs bosons are all completely
independent of one another, the only constraint arising
from the requirement that the Higgs sector contribution to $\Delta \rho$
be small.

In this situation, $\gam\gam$ collisions could play a very vital role.
Any neutral Higgs boson can be produced singly in $\gam\gam$ collisions,
including the $\ha$, whose $\gam\gam$ coupling is determined by fermion
loops (with large mass asymptotic limit of 2, compared to the $-4/3$
quoted earlier for the CP-even $\hsm$). The cross section for any given Higgs
boson depends upon the precise weighting of the different loop diagrams,
as determined by the appropriate $\beta$- and $\alpha$-dependent coupling
constant factors.\refmark\hhg\ Generally speaking the $\gam\gam$ width of all
the neutral Higgs bosons of the general two-doublet model can be substantial,
and their detection in $\gam\gam$ collisions would be possible over a large
range of parameter space.

At this point it is useful to return to the issue of whether one can
determine the quantum numbers of a neutral Higgs boson
(which we generically denote by $\hn$) that has been
detected. Here we focus on the CP quantum number. At least three
distinctly different approaches can be identified. First, there are
distributions related to the production process $\epem\rta Z^*\rta Z\hn$.
\REF\bcdkz{V. Barger, K. Cheung, A. Djouadi, B. Kniehl, P.M. Zerwas,
preprint MAD-PH-749 (1993).}
\REF\hagstong{K. Hagiwara and M.L. Stong, work in progress; see the
contribution to these proceedings by M. Stong.}
\refmark{\gwhs,\bcdkz,\hagstong}\
Second, certain  distributions of decay products, \eg\ in $\hn\rta WW\rta
f\anti f f\anti f$, are sensitive to the quantum numbers of the decaying
Higgs boson.
\REF\dkr{M.J. Duncan, G.L. Kane and W.W. Repko, \prlj{55} (1985) 579;
\npbj{272} (1986) 517.}
\REF\nelson{J.R. Dell' Aquila and C.A. Nelson, \prdj{33} (1988) 80,93,
101.}
\REF\matsuura{T. Matsuura and J. van der Bij, {\it Z. Phys.} {\bf C51}
(1991) 259.}
\REF\poiswy{H. Pois, T.J. Weiler and T.C. Yuan, \prdj{47} (1993) 3886.}
\REF\ckp{D. Chang,  W.-Y. Keung, and I. Phillips, preprint
CERN-TH-6814-93 (1993).}
\REF\sonixu{A. Soni and R.M. Xu, preprint ITP-SB-92-54 (1992).}
\REF\skjold{A. Skjold and P. Osland, preprint ISSN 0803-2696 (1993).}
\refmark{\dkr-\skjold}
Finally, the  dependence of the production rate of
a neutral $\hn$ in $\gam\gam$ collisions on the polarization of
the colliding photons can be a powerful tool in revealing the $\hn$
quantum numbers.
\REF\gg{J.F. Gunion and B. Grzadkowski, \plbj{294} (1992) 361.}
\refmark{\gg}\
In order to discuss these techniques, it will be useful to consider a
two-Higgs doublet model in a bit more detail.
In the most general case, where the Higgs potential includes CP-violating
terms, one must diagonalize a $3\times 3$ mass-squared matrix.  There will
simply be three mass eigenstates of mixed CP nature.  The coupling of the
$\hn$ to the $ZZ$ or $\wp\wm$ channel relative to the coupling of the $\hsm$
to these channels is given at tree-level by:
$${g_{WW\hn}\over g_{WW\hsm}}=u_2\sinb+u_1\cosb\eqn\wwhcoup$$
where $u_i$ ($\Sigma_{i=1,3} |u_i|^2=1$) specifies the $\hn$ mass eigenstate
in a certain basis where $i=1,2$ refer to CP-even components and $i=3$
corresponds to a CP-odd component.  Note that at tree-level the CP-odd
component does not give rise to any $WW$ coupling, and that the
net CP-even component coupling of the $\hn$ will in general be reduced
compared to the SM $\hsm$ value. The coupling of the CP-odd component of the
$\hn$ to the $WW$ channels is very small, arising from one-loop diagrams.

To illustrate the potential, but also the difficulties, of the first two
approaches mentioned above, let me consider examples of each.  In $\epem\rta
Z^*\rta Z\hn$ production, consider the distribution $d\sigma/d\cos\theta$,
where $\theta$ is the angle of the produced $Z$ in the center of mass with
respect to the direction of collision of the initial $\ep$ and $\em$. The
distribution takes the form
$${d\sigma\over d\cos\theta}\propto \cases{
{8\mz^2\over s}+{\beta^2}\sin^2\theta: &CP-even \cr
1+\cos^2\theta: &CP-odd , \cr}\eqn\proddis$$
where $\beta$ is the center-of-mass velocity of the final $Z$.
Thus, in principle one can measure this distribution and determine
the quantum numbers of the $\hn$ being produced. The difficulty with this
conclusion is best illustrated by considering a $\hn$ which is a mixture
of CP-even and CP-odd components, as described above. The crucial point is
that only the CP-even portion of the $\hn$ couples at tree-level to $ZZ$,
whereas the CP-odd component of the $\hn$ couples weakly to $ZZ$
via one-loop diagrams. Consequently, the $d\sigma/d\cos\theta$
distribution will reflect only the CP-even component of the $\hn$, even
if the $\hn$ has a fairly large CP-odd component.  The $\hn$ would have
to be almost entirely CP-odd in order for the $\cos\theta$ distribution
to deviate significantly towards the CP-odd prediction.  However, in this
case, the $\epem\rta Z^*\rta Z\hn$ production rate would be very small,
and the $\hn$ would probably not be detectable in the $Z\hn$
associated production mode in any case.  In summary, any $\hn$ {\it which is
not difficult to detect}
in the $Z\hn$ mode will automatically have a $\cos\theta$
distribution that matches the CP-even prediction, even if there is a
significant CP-odd component in the $\hn$. Unfortunately, the $Z\hn$
production rate itself cannot be used as a measure of the CP-even vs.
CP-odd component of the $\hn$; as noted earlier, even a purely CP-even
$\hn$ can have a $ZZ$ coupling that is suppressed relative
to SM strength.

Turning to decay distributions, one encounters a similar problem.
In $\hn\rta WW\rta 4~{\rm fermions}$, one can determine the angle $\phi$
between the decay planes of the two $W$'s.  One finds:
$${d\sigma\over d\phi}\propto \cases{
1+\alpha \cos\phi+\beta\cos2\phi: &CP-even \cr
1-(1/4)\cos2\phi: &CP-odd , \cr}\eqn\proddis$$
where $\alpha$ and $\beta$ depend upon the types of fermions observed and
the kinematics of the final state.
In general these two distributions are distinguishable.  However,
in analogy to the previous case, it is almost entirely the CP-even component
of the $\hn$ which will be responsible for its decays to $WW$, and the
$\phi$ distribution will thus closely match the CP-even prediction, even if
the $\hn$ has a substantial CP-odd component.
This is explicitly verified in the calculations of Ref.~\sonixu.
In order for the $\phi$ distribution to deviate significantly towards the
CP-odd prediction, the CP-even component(s) of the $\hn$ must be small.
Consequently, decays of the $\hn$ to $WW$ channels will be substantially
suppressed, and either the $b\anti b$ or $t\anti t$ channel will
dominate. In the $b\anti b$ channel, the CP-even and CP-odd
components of the $\hn$ cannot be separated. The distribution
$d\sigma/d\cos\theta^*$ (where $\theta^*$ is the $b$ angle relative to the
boost direction in the $\hn$ rest frame) is predicted to be flat,
independent of the CP nature of the $\hn$.

If $\hn\rta t\anti t$ decays are kinematically allowed, in a general
two-Higgs-doublet model they can be important or even dominant
relative to the $WW$ decay modes. Further, the CP-even and CP-odd
components of the $\hn$ would couple to the $t\anti t $ channel with
similar strength. Finally, the $t$ decays are perturbative and the lepton
spectra from the $\wpm$ decay products reflect the spin of the $t$'s.
CP-odd asymmetry observables can be constructed using the spin or lepton
momenta; a few thousand Higgs boson events might allow observation of an
effect if  $BR(\hn\rta t\anti t)$ is large.
\Ref\hemaetal{X.-G. He, J.P. Ma, and B. McKellar, preprint UM-P-92/11
(1993).}

The $\gam\gam$ coupling of the $\hn$ is another experimental observable that
`couples' with similar strength to CP-even and CP-odd components of the $\hn$,
while simultaneously revealing how much of each is present.
The CP-even component of the $\hn$ will couple to $\gam\gam$ in the fashion of
the $\hsm$, although the relative weights of $W$ and fermion loops can
easily be different. In terms of the polarization vectors $\vec e_{1,2}$ of
the two photons in the photon-photon center of mass, the coupling is
proportional to $\vec e_1\cdot \vec e_2$. The CP-odd component of the $\hn$
will also develop a $\gam\gam$ coupling at one-loop.  As noted earlier,
only fermion loops contribute. The coupling is proportional to $(\vec e_1\times
\vec e_2)_z$ (assuming the photons collide along the $z$ axis). Writing the
net coupling as $\vec e_1\cdot \vec e_2 \cale+(\vec e_1\times \vec e_2)_z
\calo$, one finds that $\cale$ and $\calo$ are naturally of similar size if
the CP-odd and CP-even components of the $\hn$ are comparable
(\ie\ $|u_1|^2+|u_2|^2\sim |u_3|^2$).

In order to reveal the appropriate
experimental observables, let us convert to a helicity basis.
One finds:
$$\eqalign{ |\calm_{++}|^2+|\calm_{--}|^2=&2(|\cale|^2+|\calo|^2)\cr
            |\calm_{++}|^2-|\calm_{--}|^2=&-4{\rm Im}\,(\cale\calo^*) \cr
            2{\rm Re}\,(\calm_{--}^*\calm_{++})=&2(|\cale|^2-|\calo|^2)\cr
            2{\rm Im}\,(\calm_{--}^*\calm_{++})=&-4{\rm
                                                Re}\,(\cale\calo^*)\,.\cr}
\eqn\helicityb$$
Each of these helicity amplitude combinations is directly observable
via production rate differences obtained by flipping or rotating the
polarizations of the colliding photons.  If the $\hn$ has both
CP-even and CP-odd components, so that $\cale$ and $\calo$ are comparable,
then the best asymmetry observable is $\cala_1\equiv
(|\calm_{++}|^2-|\calm_{--}|^2)/(|\calm_{++}|^2+|\calm_{--}|^2)$.
The term in the cross section proportional to $\cala_1$ changes sign when
both $\lam_1$ and $\lam_2$ are simultaneously flipped, and is thus
best measured by taking $<\lam_1\lam_2>$ as
near 1 as possible (which suppresses backgrounds anyway), and then
comparing the event rate for $<\lam_1>$ and $<\lam_2>$ both positive, to
that obtained when both are flipped to negative values. Experimentally
this is achieved by
simultaneously flipping the helicities of both of the initiating
back-scattered laser beams. One finds\refmark\gg\ that this asymmetry
is typically larger than 10\% and is
observable for a large range of two-doublet parameter space if CP violation
is present in the Higgs potential.

If CP violation is absent, then either
$\cale$ or $\calo$ will be zero.  To distinguish a CP-even from a CP-odd
$\hn$ will then require measurement of the asymmetry $\cala_3\equiv 2{\rm
Re}(\calm_{--}^*\calm_{++})/(|\calm_{++}|^2+|\calm_{--}|^2)$. $\cala_3$ is
$+1$ for a CP-even Higgs boson or $-1$ for a CP-odd Higgs boson. To extract
it experimentally requires that the colliding photons have transverse
polarization. For back-scattered laser beams, maximal achievable transverse
polarizations  are smaller than maximal achievable helicities, and
$\cala_3$ is more difficult to measure than $\cala_1$. Preliminary
results\Ref\kelly{J.F. Gunion and J. Kelly, work in progress.} indicate
that large (but achievable) luminosities will be required in order to
determine the sign of $\cala_3$.

\smallskip
\noindent{\bf 4. The Higgs Bosons of the Minimal Supersymmetric Model}
\smallskip

It is not possible for me to review all the attractive features of
the MSSM model here.  This subject is reviewed by G. Kane in these
proceedings. A particularly noteworthy feature of the model is the
fact that, for simple boundary conditions, grand unification in the
context of perturbative renormalization/evolution equations yields highly
satisfactory values for $\sin^2\theta_W$ and other precisely
measured electroweak parameters. In
addition, a common GUT-scale Yukawa coupling yields fermion mass ratios,
\eg\ $\mb/m_\tau$ that tend to be in close agreement with experiment.
Supersymmetry also provides a good candidate for dark matter (the lightest
neutralino), and the predicted GUT scale adequately suppresses proton decay.

The Higgs sector of the MSSM is a highly constrained two-doublet
Higgs model.  Two doublets are required by the basic structure of
supersymmetry which makes it impossible to use a single Higgs superfield
and its complex conjugate simultaneously in the
superpotential construction that is responsible for fermion masses.
(Recall that in the MSM the Higgs field yields the down quark masses,
while its complex conjugate appears in the Lagrangian term responsible
for up quark masses.) Thus, one Higgs superfield has a spin-0 component
field that gives mass to down quarks, while the spin-0 component of
the other Higgs superfield yields up quark masses.\foot{In the common
nomenclature\refmark\hhg\ this means that the two-doublet model will
be type-II, the same type as implicitly assumed in the previous section.}
Alternatively, one can also verify that two Higgs superfields are required
in order to complete the anomaly cancellations in the supersymmetric
context, where superpartners of the Higgs bosons must be present.

In the supersymmetric structure, the quartic couplings of the Higgs
fields become related to gauge couplings, and are no longer free
parameters. The quadratic mass terms are strongly constrained by
minimization conditions, and in the end only two parameters are required
to fully specify the Higgs potential.  These are normally taken to be
$\tanb$ and $\mha$, the mass of the CP-odd scalar.\foot{The MSSM
Higgs potential is such that CP violation in the Higgs sector is not
possible at tree-level.}
At tree-level, the masses and couplings of all the Higgs
bosons can be computed in terms of $\mha$ and $\tanb$. (In particular, the
mixing angle $\alpha$ is determined.)  Additional
parameters are required to determine the one-loop corrections to the Higgs
masses, which can be large if $\mt$ and $\mstop$ (the stop squark mass)
are both large.  The basic results are well-known.
\Ref\hehperspectives{For a review and references, see H.E. Haber in
\perspectives, p. 79. See also the talk by R. Hempfling at this
conference.}  The two most important points are:
\pointbegin
$\mhl\leq\mz+f(\tanb,\mt,\mstop,\ldots)$, where $\ldots$ refers to generally
less important parameters. This upper limit for $\mhl$, which is approached at
large $\mha$, increases with increasing $\tanb$, $\mt$ and/or
$\mstop$.
\point
For large $\mha$, $\mhh\sim\mhpm\sim\mha$, $\mhl$ approaches an upper
limit, and the couplings of the $\hl$ become rather SM-like.
The approach of the $\hl$ couplings to SM-like values is, however, slow
enough that important and possibly measurable deviations will be present
even for $\mha$ values above several hundred GeV.

Constraints from existing data on the MSSM Higgs sector are few.
Data from LEP-I implies that $\mha\gsim 20\gev$ and $\mhl\gsim 40\gev$.
No constraint is currently placed on $\tanb$.  The $b \rta s\gamma$
decay branching ratio limit which is such a powerful constraint in
the non-SUSY two-doublet context is considerably weaker. In the strict
supersymmetric limit, $BR(b\rta s\gam)=0$.  Not surprisingly, there
is then a large region of SUSY parameter space such that there is no
inconsistency with current upper limits.
\REF\bertolini{S. Bertolino, F. Borzumati, and A. Masiero, \npbj{294}
(1987) 321; S. Bertolini, F. Borzumati, A. Masiero, and G. Ridolfi,
\npbj{353} (1991) 591.}
\REF\bargiud{R. Barbieri and G.F. Giudice, \plbj{309} (1993) 86.}
\REF\lopez{J. Lopez, D. Nanopoulos, and G. Park, preprint CTP-TAMU-16-93
(1993).}
\REF\oshimo{N. Oshimo, preprint IFM 12/92 (1992).}
\refmark{\bertolini-\oshimo}

\FIG\lepii{}
\topinsert
\vbox{\phantom{0}\vskip 5.0in
\phantom{0}
\vskip .5in
\hskip -110pt
\special{ insert user$1:[jfgucd.rcsusyhiggs]erice_92_contour_lepii.ps}
\vskip -.15in }
\centerline{\vbox{\hsize=12.4cm
\Tenpoint
\baselineskip=12pt
\noindent
Figure~\lepii: The boundaries in $\mt$ and $\mstop$ parameter space, beyond
which the $\hl$ cannot be detected in $Z^*\rta Z\hl$ at LEP-II, are plotted.
We assume that
$L=500\pbi$ and require 100 total $\hl Z$ events (before efficiencies).
Each curve corresponds to a different choice for the machine energy
and $\tanb$, as indicated by the $(\sqrt s,\tanb)$ labellings.
$\mha$ has been chosen to be 1 TeV, so that $\mhl$ is near its upper limit.
We have neglected squark mixing. The region where discovery is possible
lies to the left and below the boundary curves.
}}
\endinsert

The first issue of importance is to specify the range of parameter space
for which LEP-II will be able to detect the $\hl$.
This has received much attention in the literature, and it is not
possible to cite all references here.  The graph I present, Fig.~\lepii, is
taken from
\REF\gunperspectives{J.F. Gunion, in \perspectives, p. 179.}
Ref.~\gunperspectives. A more experimentally oriented discussion is
given in Ref.~\janotstudy.
Figure~\lepii\ shows the discovery boundaries
in $\mt$--$\mstop$ parameter space beyond which $\hl$ detection
(at large $\mha$) would be impossible, as a function of the $(\sqrt
s,\tanb)$ values chosen.  The following observations are perhaps useful:
\pointbegin
If $\mt=180\gev$, $\sqrt s\gsim 240\gev$ is required for detection to be
possible at large $\mstop$ and large $\tanb$.
\point
If $\mt\lsim140\gev$, $\sqrt s\gsim 200\gev$ yields good coverage,
but $\sqrt s\lsim 190\gev$ would not. For instance, if $\mstop\sim
350\gev$, a value representative of that found in some grand unification
scenarios, $\sqrt s\sim 200\gev$ would cover all $\mt$ and $\tanb$
possibilities (within reason), whereas $\sqrt s\sim 190\gev$ would leave a
region at large $\tanb$ in which $\hl$ detection would not be possible.

\FIG\nlc{}
\topinsert
\vbox{\phantom{0}\vskip 5.5in
\phantom{0}
\vskip .5in
\hskip -100pt
\special{ insert user$1:[jfgucd.rcsusyhiggs]perspectives_nlc.ps}
\vskip -.15in }
\centerline{\vbox{\hsize=12.4cm
\Tenpoint
\baselineskip=12pt
\noindent
Figure~\nlc:
Event number contours for an NLC with $\sqrt s=500\gev$ and $L=10\fbi$.
We have taken $\mt=150\gev$, $\mstop=1\tev$ and have neglected
squark mixing. We display contours for 30 and 100 total events
(no detection efficiencies included) for
the six most important processes at an NLC:
$Z^*\rta Z\hl,Z\hh,
\hl\ha,\hh\ha$, and $WW\rta \hl,\hh$. Generally speaking, 100 events
ensure that the reaction can be observed.}}
\endinsert

If the $\hl$ is not found at LEP-II because the machine energy is
not pushed to the required level, then one
must rely on future colliders to probe the MSSM Higgs sector.  An
$\epem$ linear collider could prove to be an ideal machine for this purpose.
\REF\brignoleetal{A. Brignole, J. Ellis, J.F. Gunion, M. Guzzo, F. Olness,
G. Ridolfi, L. Roszkowski and F. Zwirner, in \munichnlc, p. 613.}
\REF\dkzepemstudy{A. Djouadi, J. Kalinowski, P.M. Zerwas, in \munichnlc,
p. 107.}
\REF\yamadastudy{A. Yamada, {\it Mod. Phys. Lett.} {\bf A7} (1992) 2877;
and contribution to these proceedings.}
A number of recent studies have appeared from both a theoretical
\refmark{\brignoleetal-\yamadastudy}\ and an experimental
perspective.\refmark{\janotstudy}  See also Ref.~\gunperspectives.
Certainly, any $\epem$ collider with $\sqrt s \gsim 300 \gev$ is guaranteed
to find the $\hl$.  At $\sqrt s=500\gev$,
if $\mha$ is large enough that the $\hl$
has SM-like couplings, then the $WW\rta\hl$ fusion process yields the
largest rate for $\hl$ production. But $Z^*\rta Z\hl$ also yields an
entirely adequate event rate for $\hl$ discovery. In Fig.~\nlc, the 30 and
100 event number contours for these and four other basic production
processes are displayed in $\mha$--$\tanb$ parameter space. Typically, 100
events (before cuts and branching ratios) should be adequate for
observation of any given process.  We see that it is only in the large
$\tanb$,  $\mha\lsim 100\gev$ corner of parameter space (where the $WW\hl$
coupling -- proportional to $\sin^2(\beta-\alpha)$ -- is small) that one
must turn to the alternative, but perfectly adequate $Z^*\rta \hl\ha$
production mode. Detection of the $\epem\rta \hl t\anti t$ process is
likely to be possible whenever the $\hl$ has roughly SM-like couplings,
\ie\ when $\mhl$ is near its upper limit.\refmark\djouadittbar\

The most important limitation of a $\epem$ collider is also apparent
from Fig.~\nlc.  We see that if $\mha\gsim 100\gev$, then the only
possible means for detecting the $\hh$ and $\ha$ is via the pair
production process, $Z^*\rta\hh\ha$.  However, the parameter range
for which this process has adequate event rate is limited
by the machine energy to $\mha\sim\mhh\lsim \sqrt s/2-30\gev$ (recall that
$\mhh\sim\mha$ at large $\mha$).  At $\sqrt s=500\gev$, this
means $\mha\lsim 210\gev$. Meanwhile, $\epem\rta \hp\hm$ is also limited to
$\mhpm\sim\mha\lsim 210-230\gev$.\refmark\komamiya\  Thus, it could happen
that only a rather SM-like $\hl$ is detected in $\epem$ collisions at the
linear collider, and none of the other Higgs bosons are observed.

\FIG\bslaser{}
\topinsert
\vbox{\phantom{0}\vskip 5.0in
\phantom{0}
\vskip .5in
\hskip -10pt
\special{ insert scr:hawaii_bslaser.ps}
\vskip -1.65in }
\centerline{\vbox{\hsize=12.4cm
\Tenpoint
\baselineskip=12pt
\noindent
Figure~\bslaser: Number of events as a function of Higgs mass
in various channels for $\gam\gam\rta \hh$ and $\gam\gam\rta \ha$
at $\mt=150\gev$.  Results for $\tanb=2$ and $\tanb=20$ are shown.
For the $b\anti b$ and $t\anti t$ channels, the continuum $\gam\gam\rta
b\anti b$ and $t\anti t$ background rates are shown for the most
optimistic possible experimental resolution.  Backgrounds in
the $\hh\rta \hl\hl$ and $\ha\rta Z\hl$ are negligible unless
$\mhl\sim\mw$, in which case $b$ tagging would be needed to eliminate
the large $\gam\gam\rta \wp\wm$ continuum background process.
These plots are machine energy independent; $\sqrt s$
would have to be about 20\% higher than the Higgs mass.}}
\endinsert

However, $\gam\gam$ collisions using back-scattered laser beams might allow
discovery of the $\hh$ and/or $\ha$ up to higher masses.\refmark\ghgamgam\
And, detection of the $\hl$ in $\gam\gam$ collisions is
relatively certain to be possible.\foot{Of course, since charged Higgs
bosons can only be pair produced in $\gam\gam$ collisions, $\epem$
collisions will yield the greatest kinematical reach in $\mhpm$.
For a study of the $\gam\gam\rta \hp\hm$ process see
\REF\dbcct{D. Bowser-Chao, K. Cheung, and S. Thomas,  preprint
NUHEP-TH-93-7 (1993).} Ref.~\dbcct.}
Observation of any of the three neutral Higgs bosons
would constitute a measurement of the $\gam\gam$~Higgs coupling, which in
principle is sensitive to loops involving other charged supersymmetric
particles such as squarks and charginos. Fig.~\bslaser\ illustrates the
discovery potential for the $\hh$ and $\ha$ at $\mt=150\gev$ in various
final state channels. (Superpartner masses have been taken to be large and
machine energy is assumed to be about 20\% higher than the Higgs mass.)
Particularly interesting channels at moderate $\tanb$ and below $t\anti t$
threshold are $\hh\rta\hl\hl$ (leading to a final state containing 4 $b$
quarks) and $\ha\rta Z\hl$.  These channels are virtually background free
unless $\mhl\sim\mw$, in which case the large $\gam\gam\rta \wp\wm$
continuum background would have to be eliminated by $b$-tagging. Above
$t\anti t$ threshold, $\hh,\ha\rta t\anti t$ decays dominate (at moderate
$\tanb$). We see that the event rate is high and that the $\gam\gam\rta
t\anti t$ continuum background is of the same general size as the signal
rate. Discovery of the $\ha$ and $\hh$ up to roughly $0.8\sqrt s$ would be
possible.

For large $\tanb$, it is necessary to look for the $\ha$ and $\hh$ in the
$b\anti b$ final state. For the effective integrated luminosity chosen,
$L=20\fbi$, Fig.~\bslaser\ shows that detection will be difficult except at
low masses, in particular masses such that $Z^*\rta \hh\ha$ would be
observable in $\epem$ collisions. However, it is technically feasible
(although quite power intensive) to run the $\gam\gam$ collider at very
high instantaneous luminosity\Ref\borden{D. Borden, private communication.
See also the talk by V. Telnov, these proceedings.}\ such that accumulated
effective luminosities as high as $200\fbi$ can be considered. In this
case, detection of the $\ha$ and $\hh$ in the $b\anti b$ channel
should be possible for masses $\lsim0.8\sqrt s$.

A particularly interesting question is the extent to which the
$\gam\gam$ widths (that would be measured by detection of the $\hl$,
$\hh$, and/or $\ha$ in $\gam\gam$ collisions) depend upon the
the SUSY context and/or superpartner masses. Some exploration of this
\REF\kileng{B. Kileng, preprint ISSN 0803-2696 (1993).}
issue has appeared in Refs.~\ghgamgam\ and \kileng.
Potentially, these widths are sensitive to  loops containing heavy charged
particles.  However, it must be recalled that supersymmetry decouples when
the SUSY scale is large. (In particular, superpartner masses come primarily
from soft SUSY-breaking terms in the Lagrangian and not from the Higgs
field vacuum expectation value(s).)
Several cases are illustrated in Fig.~\widthratios.
First, suppose the lightest CP-even Higgs boson has been
discovered, but that no experimental evidence for either the  heavier Higgs
bosons or any supersymmetric particles has been found. Could a measurement
of the $\hl\gam\gam$ coupling provide indirect evidence for physics beyond
the SM? Figure~\widthratios\ shows that if the MSSM parameters are
chosen such that all new particles beyond the SM are too heavy to be
produced (technically we take $M=-\mu=300\gev$ for the charginos
and a common squark/slepton diagonal mass of 300 GeV), then the
deviation of $\Gamma(\hl\rta\gam\gam)$ from the corresponding SM value for
the $\hsm$ is less than 15\%. This is because of decoupling; as the SUSY
breaking scale and the scale of the heavier Higgs bosons become large, all
couplings of the $\hl$ approach their SM values and the squark and
chargino loops become negligible. Even with the MSSM parameters chosen
such that the supersymmetric partners lie only just beyond the reach of a
$\sqrt s=500\gev$ $\epem$ collider, it will not be easy to distinguish the
$\hl\rta\gam\gam$ decay width from that of the $\hsm$. Ref.~\bbcnew\
claims measurement accuracies for the $\gam\gam$ width
of a SM-like Higgs boson of order 10\%, \ie\
just on the borderline of what is required.

On the other hand, suppose that the $\hh$ or $\ha$ is light enough to be
seen in $\gam\gam$ collisions.  In this case, a measurement of its
$\gam\gam$ coupling can provide useful information
on the spectrum of charged supersymmetric particles (even if the latter are
too heavy to be directly produced). Fig.~\widthratios\ provides two
examples in the case of the $\hh$. Large alterations in the $\gam\gam$
width occur if either (or both) the chargino mass scale or the squark mass
scale is taken to be significantly below 1 TeV. In the case of the $\ha$,
squark loop contributions to the $\ha\gam\gam$ coupling are absent.
However, the sensitivity of this coupling to the chargino loops is similar
to that of the $\hh\gam\gam$ coupling.

Of course, if only one light Higgs boson is discovered, its $\gam\gam$
coupling is only one among many that might reveal that the Higgs boson is
the $\hl$ of MSSM rather than the $\hsm$. Hildreth\refmark\hildreth\ shows
that $b$ tagging can separate the various $\hl$ decay channels sufficiently
that the $\hl$--$\hsm$ distinction can be made if $\tanb\gsim 6$.  The
coupling that deviates most from the SM value when $\tanb$ is large is the
$WW$ coupling, but the (harder to measure) $c\anti c+ g g$ channel ($c\anti
c$ and $gg$ are combined since they have similar displaced vertex rates)
also has large deviations from SM expectations.
In another contribution to this conference, Kurihara
\Ref\kurmiy{J. Fujimoto, Y. Shimizu, T. Munehisa, N. Nakazawa,
and Y. Kurihara, as presented by Y. Kurihara, these proceedings.}
has claimed that at high integrated luminosity ($L\sim 150\fbi$) the
measurement of $\sigma(\epem\rta Z\hl)\times BR(\hl\rta b\anti b)$ can be
made with such precision (2\%) that the distinction between the $\hl$ and
the $\hsm$ would be possible over most of parameter space (for $\mha\lsim
600\gev$). However, there are systematic uncertainties, such as our
inability to precisely determine the $b$ quark mass.  Further study of the
influence of such systematic problems appears to be necessary before
drawing any firm conclusions.

Once again, it is appropriate to compare the ability of an $\epem$
collider to that of the SSC/LHC hadron colliders to probe the MSSM Higgs
sector. For reviews of the results, see Ref.~\gunperspectives\ and the
talk by A. Rubbia in these proceedings. If only Higgs$\rta\gam\gam$
(possibly in association with $t\anti t$ production), Higgs$\rta
ZZ^{(*)}\rta 4\ell$ and $t\rta \hp b$ channels are deemed sufficiently
clean to be viable at a hadron collider,
then it is possible for there to be a window in
$\mha$--$\tanb$ parameter space for which no MSSM Higgs boson is observed
at either LEP-200 or at the SSC/LHC. This parameter space hole, located in
the range $110\lsim\mha\lsim 170\gev$ and $\tanb\gsim 7-10$, only arises if
$\mt$ is in the vicinity of $130-170\gev$, if $\mstop\sim 1\tev$, and if no
other detection modes are viable.  If $\mt$ is either smaller ($\sim
100\gev$) or larger ($\sim 200\gev$), or if $\mstop$ is smaller ($\lsim
400\gev$) then the hole disappears even without considering additional
modes.

One additional mode that has been investigated is $\hh,\ha\rta \tauptaum$.
It becomes viable at large $\tanb$, where the $gg\rta b\anti b+\ha,\hh$
production rates are greatly enhanced and $b\anti b$ decays have $\sim
90\%$ branching ratio.  In the L3P simulation study reviewed by Rubbia,
$\ha$ and $\hh$ detection in this mode is viable for all $\mha\gsim
100\gev$ and $\tanb\gsim 7$. In particular, this mode is unique in that
rather heavy $\ha$ and $\hh$ Higgs bosons can be seen if $\tanb$ is
larger enough.

A second mode that has just recently received attention is associated
Higgs+$t\anti t$ production with Higgs$\rta b\anti b$.
\REF\dgvii{J. Dai, J.F. Gunion and R. Vega, preprint UCD-93-20 (1993).}
Based on the analysis of the SM Higgs boson,\refmark\dgvi\ after correcting
for branching ratio and production rate differences, Ref.~\dgvii\
concludes that if $\mt\sim 150\gev$, then detection of the $\hl$
in this mode will be possible for any $\mha\gsim 110\gev$; $\hh$
detection in this mode is possible for $\mha\gsim 50\gev$ up to
the lower limit in $\mha$ for $\hl$ detection.
\foot{There is a crossover in the vicinity of $\mha\sim 110\gev$
where the $\hl$ and $\hh$ interchange roles as being SM-like.}
That is, either the $\hh$ or the $\hl$ can be detected in this way for
$\mha\gsim 50\gev$. This completely closes the parameter space hole. In
fact, by combining just the $t\rta \hp b$ detection mode and this $t\anti t
b\anti b$ final state mode, detection of at least one MSSM Higgs boson is
guaranteed to be possible at the SSC/LHC alone.
At $\mt\sim 200\gev$, the parameter space region for which the $t\anti
tb\anti b$ mode is viable (for either $\hl$ or $\hh$) is smaller; but this is
simply correlated with the fact that other decay modes, most notably the
$ZZ^{(*)}\rta 4\ell$ mode, of the $\hh$ and (now rather heavy) $\hl$ acquire
larger branching ratios, and become viable over a large range of
parameter space. Of course, it should be noted that this $t\anti t b\anti
b$ mode is not viable for actually observing a heavy $\hh$ or heavy $\ha$;
at large $\tanb$ the $t\anti t \ha$ and $t\anti t \hh$ production
processes are suppressed, while at small $\tanb$, the $\ha$ and $\hh$ decays
would be dominated by the $t\anti t$ final state.

To reiterate, combining all modes, the SSC/LHC alone will detect
at least one and most probably several of the MSSM Higgs bosons. In
addition, it is not unlikely that the $\hl$ can be detected in
all three of its most crucial decay channels, $ZZ^*$, $b\anti b$, and
$\gam\gam$, simultaneously. Further, detection of rather heavy  $\hh$ and
$\ha$ will be possible if $\tanb$ is large.  An $\epem$ collider would
undoubtedly do a better job of determining all the $\hl$ couplings,
but the $\ha$,  $\hh$  and $\hpm$ could be beyond its kinematical reach.

The above discussion does not take into account possible supersymmetric
decay channels for the MSSM Higgs bosons.  When allowed, $\chitil\chitil$
(where $\chitil$ represents a chargino or neutralino)
decay modes of the MSSM Higgs are substantial, and often dominant.
Some work on this subject at the SSC/LHC has appeared in
\REF\baersusy{H. Baer, M. Bisset, D. Dicus, C. Kao, and X. Tata, \prdj{47}
(1993) 1062.}
Refs.~\baersusy\ and \gunperspectives.
The kinematic reach of a $\sqrt s\sim500\gev$ $\epem$ collider is such that
these modes are  not likely to be relevant unless the lightest chargino is
seen at LEP-200.\refmark{\dkzepemstudy}\

Of course, in more specific grand unification, renormalization group
scenarios, rather restricted predictions for the SUSY parameters emerge.
This very active field is reviewed in these proceedings in the
contributions by G. Kane, V. Barger, S. Pokorski, and L. Roszkowski.
Very often the resulting $\mt$ value is in the $130-180\gev$ range,
and $\tanb$ lies between 2 and 10. Squark masses are often of rather
moderate size, implying that the $\hl$ is relatively light, $\mhl\lsim
110\gev$.
\foot{Of course, $\mstop$ cannot be light if $\tanb$ is near 1 without
violating current limits from LEP-I on a light Higgs boson.}
But, the $\hh$ and $\ha$ are generally in the mass range above
$220\gev$ (\ie\ just beyond the reach of $\epem$ collisions at a $\sqrt
s=500\gev$ collider).  Neutralinos and charginos are light enough
that SUSY decay modes of the $\ha$ and $\hh$ would be important.
Generically speaking, the lightest chargino is typically light enough to be
seen at LEP-200. If SUSY really lives at this low a mass scale, we shall
soon know it! However, it should be emphasized that all these
renormalization group/GUT investigations make rather specific assumptions
(\eg\ limits on fine-tuning, bottom-$\tau$ Yukawa unification, \etc).  It
could still be that nature chooses a SUSY scale nearer 1 TeV.

What if one goes beyond the MSSM? As reviewed by G. Kane for these
proceedings, in supersymmetric models there will always be a light Higgs
boson with mass below roughly 150 GeV, assuming the absence of additional
new physics between the scale of supersymmetry breaking and $\sim 10^{16}\gev$.
(In fact, G. Kane argues for a much
lower upper limit in most cases.) Thus, the lightest Higgs boson of the
model will always be kinematically accessible at a $\sqrt s \sim 500\gev$
$\epem$ machine.  However, there are often quite a few CP-even
Higgs eigenstates, and it could happen that these share the $WW$
coupling in such a way that none have a large production event rate.
This was investigated recently
\REF\kimstudy{B. Kim,  contribution to these proceedings.}
in Ref.~\kimstudy, using the model in which there is one additional
Higgs singlet field.
\Ref\ellisetalsinglet{J. Ellis, J. Gunion, H. Haber, L. Roszkowski,
F. Zwirner, \prdj{39} (1989) 844.}\
For a significant range of acceptable renormalization group
solutions, all the CP-even Higgs bosons are light (in agreement
with Ref.~\ellisetalsinglet) but none would have adequate $WW$
coupling for discovery.

\smallskip
\noindent{\bf 5. Conclusions}
\smallskip

Overall, it is clear that a $\sqrt s\gsim 500\gev$ $\epem$ collider
has great potential for Higgs boson discovery and, especially,
detailed study. In both the SM and the MSSM model,
light Higgs boson(s) should be regarded as likely, and any such light
Higgs is almost certain to be discoverable at an $\epem$ machine.
The principle limitation of an $\epem$ collider relative to the SSC/LHC
hadron machines is kinematic. But, the couplings and branching ratios of
any Higgs boson that can be detected at the $\epem$ collider can be more
accurately investigated in detail than can those of a Higgs boson found at
the SSC/LHC. The importance of implementing the back-scattered laser beam
technique for probing the Higgs sector cannot be
overemphasized.  Not only could it lead to increased kinematic reach
for the $\epem$ collider, but it will allow a measurement of the
very interesting (as a probe of new physics) $\gam\gam$ coupling
of any Higgs boson seen, and may allow a determination of its CP
properties.

\smallskip\noindent{\bf 6. Acknowledgements} \smallskip
This review has been supported in part by Department of Energy
grant \#DE-FG03-91ER40674
and by Texas National Research Laboratory grant \#RGFY93-330.
I would like to thank the Aspen Center for Physics for support
during its preparation.  I also gratefully acknowledge
the contributions of my many collaborators to the content of this
report.

\smallskip
\refout
\end